\documentclass[final]{aipproc}

\layoutstyle{8x11double}
\usepackage{indentfirst}
\begin{document}

\title{Fostering Computational Thinking In Introductory Mechanics}


\author{Marcos D. Caballero}{
  address={School of Physics, Georgia Institute of Technology, Atlanta, GA 30332},
  altaddress={Current Address: Department of Physics, University of Colorado, Boulder, CO 80309}
}

\author{Matthew A. Kohlmyer}{
  address={Department of Physics, North Carolina State University, Raleigh, NC 27695},
  altaddress={Current Address: Advanced Instructional Systems, Inc., Raleigh, NC 27606}
}

\author{Michael F. Schatz}{
  address={School of Physics, Georgia Institute of Technology, Atlanta, GA 30332}
}

\begin{abstract}
Students taking introductory physics are rarely exposed to computational modeling. In a one-semester large lecture introductory calculus-based mechanics course at Georgia Tech, students learned to solve physics problems using the VPython programming environment. During the term 1357 students in this course solved a suite of fourteen computational modeling homework questions delivered using an online commercial course management system. Their proficiency with computational modeling was evaluated in a proctored environment using a novel central force problem. The majority of students (60.4\%) successfully completed the evaluation. Analysis of erroneous student-submitted programs indicated that a small set of student errors explained why most programs failed. We discuss the design and implementation of the computational modeling homework and evaluation, the results from the evaluation and the implications for instruction in computational modeling in introductory STEM courses.
\end{abstract}

\maketitle


\section{\label{sec:vpintro}Introduction}

Computation (the use of the computer to solve numerically, simulate or visualize a physical problem) has revolutionized scientific research and engineering practice. 
In science and engineering, computation is considered to be as important as theory and experiment \cite{siamstatementweb}. 
Yet, in sharp contrast, most introductory courses fail to introduce students to computation's problem solving powers. 
To provide undergraduate students opportunities for developing modeling and numerical analysis skills, we have introduced computational homework in a large enrollment introductory calculus-based mechanics course at the Georgia Institute of Technology.  

We have built upon previous work that introduces computation into introductory physics laboratories \cite{computajp,beichner2010labs}.  We use a freely available programming environment (VPython) and leverage our experience with the Matter \& Interactions (M\&I) textbook \cite{vpythonWebsite, mandi1}.
M\&I uses the open source VPython environment to teach computation because VPython has a number of helpful features that enable novice programmers to 
construct high-quality three-dimensional simulations easily. 
Thus, using VPython, students with no previous programming experience are able
to view, alter or construct computational models of force and motion problems 
directly (i.e., no physics is hidden inside a ``black box'' simulation).
In the traditional implementation of M\&I, the practice of constructing computational models is limited to the laboratory.   We use the laboratory experiences as the starting point for computational homework.  It is worth emphasizing 
that, while our implementation builds on our M\&I experience, the use of computational homework, in our view, need not be restricted only to courses that use the M\&I curriculum.

\section{Design And Implementation Of Computational Homework}\label{sec:vpdeploy}

Our design philosophy starts with the idea that students should learn 
computation by altering their own lab-developed programs to solve slightly modified problems outside of lab. 
This design philosophy emulates the practice of research scientists, who write a program to solve a problem and then, alter that program to solve different problems.  In the same spirit, we envisioned developing computational activities that would start with students working, under TA guidance, in the laboratory to develop a program that solves a problem. Students would then use that program individually to solve a different problem on their homework by making any modifications that were necessary.  The range of problems that becomes accessible to students who have learned computation is large and diverse; we chose to focus our efforts on teaching students to apply Newton's Second Law iteratively to predict motion.

\begin{figure}
\includegraphics[width=0.45\textwidth, clip, trim=0mm 0mm 0mm 0mm]{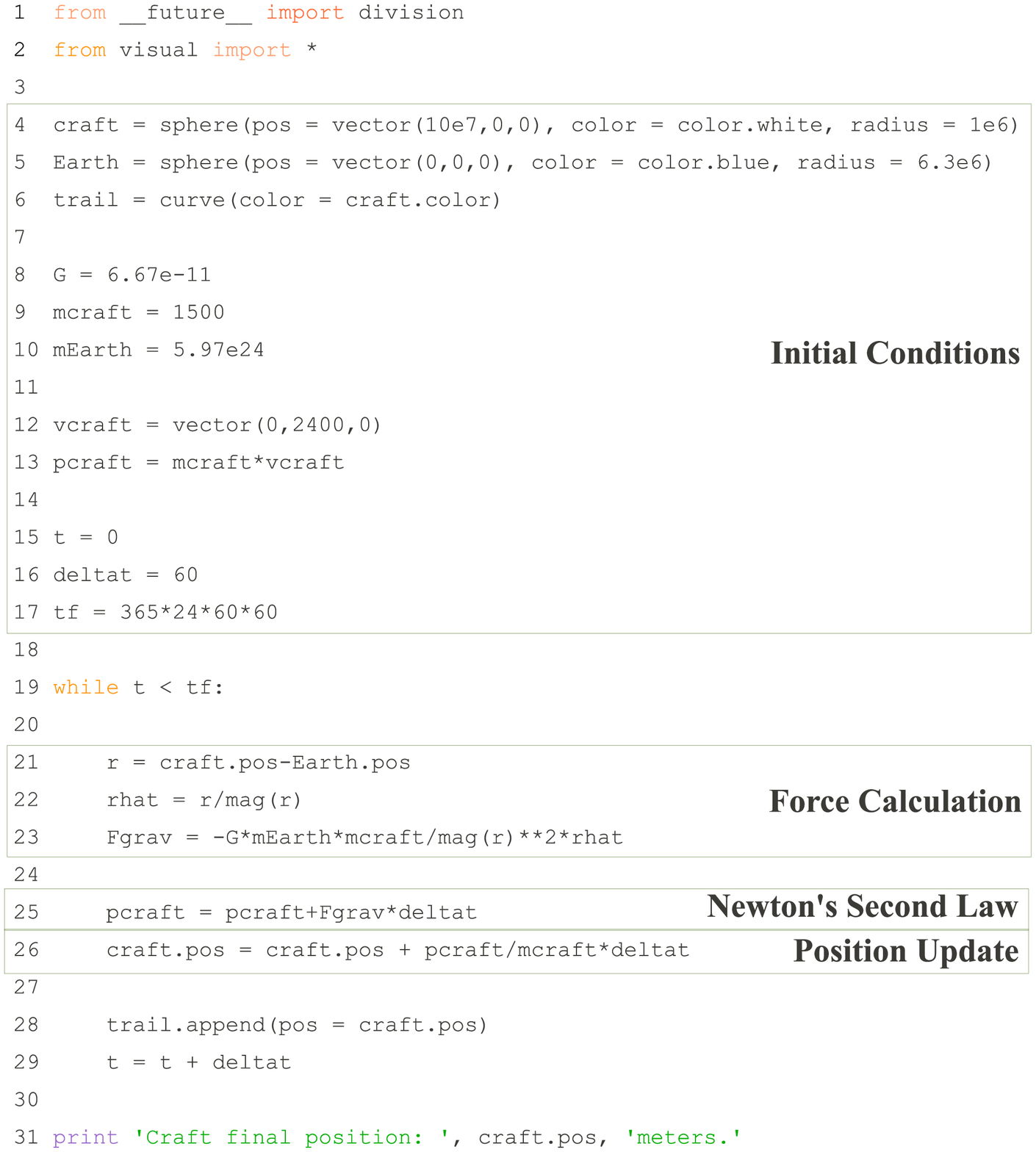}
\caption{[Color] - Under the guidance of their TAs, students wrote the VPython program above in the laboratory. This program modeled the motion of a craft (size exaggerated for visualization) orbiting the Earth over the course of one "virtual" year. To construct this model, students must create the objects and assign their positions and sizes (lines 4--6), identify and assign the other given values and relevant initial conditions (lines 8-10, 12--13 and 15--17), calculate the net force acting on the object of interest appropriately (lines 21--23) and update the momentum and position of this object in each time step (lines 25--26).}\label{fig:hwsamplecode}
\end{figure}

As a concrete example of our design, we show a mid-semester laboratory activity and homework problem in which students modeled a two-body gravitational interaction. Students develop a VPython program in the laboratory that models the motion of a craft as it orbits the Earth (Fig. \ref{fig:hwsamplecode}).
While in the lab, students could, at any time, seek help from TAs with the model.  Later, outside of lab, students make a number of modifications to this program to solve a new problem on their homework.
For example, students could be asked to alter their initial conditions to predict the position and velocity of the craft after some integration time.

Computational homework problems are deployed using the WebAssign course management system, which facilitated the weekly grading of students' solutions. 
To create the homework problem, we numerically integrated several hundred initial conditions for the physical system of interest and stored the solutions, including final quantitative and qualitative results. 
Each student was assigned a random set of initial conditions that ensured each student received a unique realization with high probability. 
On these weekly assignments, only students' final results (numeric answers, etc.) were graded; their code was uploaded for verification purposes, but not graded.
Computational homework problems were generally completed in the week that followed the associated laboratory activity.

To help students to debug their programs, each assignment contained a {\it test case} -- an initial problem was posed for which the full solution (i.e., the results from the numerical integration) was given.
This test case ensured that a student's program worked properly and helped to instill confidence in students who might otherwise have been uncomfortable writing VPython programs without the help of their group members or TAs.
After a student checked her program against the {\it test case}, she completed the {\it grading case}, a problem without a given solution. 

\section{Evaluating Computational Modeling Skills}\label{sec:eval}

Students solved both analytic and computational modeling homework problems 
with equal success (Analytic HW 84.6\% vs Computational HW 85.8\%) \cite{conover_nonpara}. 
However, this result does not indicate what fraction of students were able to solve these computational homework problems without assistance. 
While randomizing initial conditions between each student's realization ensured that students' solutions differed with high probability, working programs could be distributed easily from student to student by email.

We delivered a proctored laboratory assignment during the final lab of 
the semester
to evaluate students' computational skills on an individual basis. 
Students received a partially completed program that created two objects (one low-mass and one high-mass), initialized some constants and defined the numerical integration loop structure.
We aimed to evaluate students' engagement in the modeling process by contextualizing a physics problem into a programming task. 
Furthermore, certain programming skills were being assessed, namely, students' abilities to identify and assign variables and implement the numerical integration algorithm. 
The assignment was delivered using WebAssign in a timed mode (30 minutes), and TAs were not permitted to help students debug their programs.
The format of the assignment was identical to students' final two homework problems; students were given a test case to check their solution before solving the grading case. 

For this assignment, students modeled the motion of the low-mass object as it interacted with the high-mass object through a central force. 
The nature of the force (attractive or repulsive) and its distance dependence ($r^n$) were randomized on a per student basis. 
We also randomized some of the variable names in the partially completed program to hinder copying. 
After adding and modifying the necessary program statements, students ran their program and reported the final location and velocity of the low-mass object.
During the assignment, students did not receive feedback from the WebAssign system about the correctness of their solution, but they were given three attempts to enter their answers.
Similar to students' online homework, only the final numerical answer was graded.

Overall, we found that, on average over three semesters, 60\% of students 
completed the proctored assignments with complete success.  To determine 
exactly what challenges the remaining 40\% faced while completing this 
assignment, we reviewed the program of each student who failed to model 
the grading case.

\begin{center}
\begin{table}[t]
\caption{Incorrectly written programs were subjected to an analysis using a set of codes developed from common student mistakes. The codes focused on three procedural areas: {\it using the correct given values} (IC), {\it implementing the force calculation} (FC) and {\it updating with the Newton's Second Law} (SL). We reviewed each of the incorrectly written student programs for each of the features listed below.\label{tab:vpcodes}}
\begin{tabular}{ p{0.05\textwidth} p{0.35\textwidth} }\hline
\multicolumn{2}{l}{\bf Using the correct given values (IC)} \\
IC1 & Used all correct given values from grading case\\
IC2 & Used all correct given values from test case\\
IC3 & Used the correct integration time from either the grading case or test case\\
IC4 & Used mixed initial conditions\\
IC5 & Exponent confusion with $k$ (interaction constant)\\\hline

\multicolumn{2}{l}{\bf Implementing the force calculation (FC)} \\
FC1 & Force calculation was correct\\
FC2 & Force calculation was incorrect but the calculation procedure was evident\\
FC3 & Attempted to raise separation vector to a power\\
FC4 & Direction of the force was reversed\\
FC5 & Other force direction confusion\\\hline

\multicolumn{2}{l}{\bf Updating with Newton's Second Law (SL)} \\
SL1 & Newton's Second Law (N2) was correct\\
SL2 & Incorrect N2 but in an update form\\
SL3 & Incorrect N2 attempted update with scalar force\\
SL4 & Created new variable for $\vec{p}_f$\\\hline
\end{tabular}
\end{table}
\end{center}

\section{Systematically Unfolding Common Errors}\label{sec:vpcat}

Students must perform several tasks to successfully write and execute the program for the proctored assignment. 
Students must interpret the problem statement; that is, they must contextualize a word problem into a programming task. 
They must review the partially completed program and identify the variables to update. 
Students need to apply their knowledge of predicting motion using VPython to the problem. 
They must identify that the force is non-constant and then write the appropriate programming statements to calculate the vector force. 
Students need to then complete the motion prediction routine by writing a statement to update the momentum of the low-mass object. 

Determining where students encountered difficulties with these tasks might help explain how students learn this algorithmic approach to use Newton's Second Law to predict motion.  
Because we reviewed students' programs after they were written, we are unable to comment directly on students' challenges with contextualizing the problem. 
Our work was limited to analyzing students' procedural efforts (i.e., identifying variables and implementing the numerical integration algorithm). 
However, some information about students' thoughts and actions could be inferred from this analysis.

Using an iterative-design approach, we developed a set of binary (affirmative/negative) codes to check which tasks students performed correctly and which errors they made. 
An initial review of students' uploaded programs yielded the mistakes that were made most often. 
These common mistakes formed the basis for the codes. The codes were developed empirically and several iterations were made before they were finalized. 
Two raters tested the codes by coding a single section of student submitted programs ($N = 45$). 
The raters resolved their differences which further explicated the codes and then recoded the section. 
The final codes (Table \ref{tab:vpcodes}) were used by both raters independently to code the remaining sections ($N = 324$). 
The final codes had high inter-rater reliability; both raters agreed on 91\% of the codes. 

We classified the codes into one of three procedural areas: {\it using the correct given values} (IC), {\it implementing the force calculation} (FC) and {\it updating with the Newton's Second Law} (SL). 
These areas were congruent with the broad range of difficulties which students exhibited through their erroneous programs. 

The pattern of common errors were investigated using cluster analysis to determine similarities among students' erroneous code.
Cluster analysis is particularly well suited for this application because it characterizes patterns in complex data sets \cite{clustereveritt}.
We used cluster analysis to determine the major features in students' incorrect programs which were responsible for their failure.

\begin{center}
\begin{table}[t]
\caption{Cluster analysis revealed that 80\% of students with erroneous code fell into one of six 
error clusters.
Most students had difficulties with calculating the force in the numerical integration loop.
Some students made mistakes in identifying and assigning their initial conditions while
few students had trouble invoking Newton's Second Law. \label{tab:error}}
\begin{tabular}{p{0.39\textwidth}|r}
\multicolumn{1}{l}{\bf Dominant Error} & \multicolumn{1}{c}{{\bf \%}}\\\hline
Stuck on Test Case; Error in Force Calculation & 23.8\\
Constructed Working Code; Error in Initial Conditions & 19.8\\
Used Net Force Magnitude of Update & 13.3\\
Stuck on Grading Case; Error in Force Calculation & 10.8\\
Raised Separation Vector to Power & 7.6\\
Force Calculation Outside Integration Loop (no update) & 7.1\\\hline
\end{tabular}
\end{table}
\end{center}

More than 80\% of students appeared in only 6 clusters (Table \ref{tab:error}). 
Many students (24\%) tended to remain stuck on the test case due to an error in their force calculation. 
These students worked diligently to solve the test case but were unable to do so. As a result, they did not proceed to the grading case. 
About 20\% of students made mistakes while replacing the given values and initial conditions.
Most students in this cluster were able to construct a working albeit incorrect program. 
Given their unfamiliarity with general central force interactions, these students might have believed their solutions were correct. 
Some students (13\%) computed the magnitude of the net force and attempted to update the vector momentum with this scalar force. 
This mathematically impossible operation would have raised a VPython error. Students in this cluster were unable to parse this error into any useful information.
Nearly 11\% of students tended to make errors in the force calculation, but worked with the grading case. 
These students might have started working with the test case, but we think it is more likely that they jumped right into working with the grading case because the dominant error appears in their force calculations.
A small number of students (8\%) tended to have the sole error of raising the separation vector to a power. 
Again, this mathematically impossible operation would have raised a VPython error but these students were unable to parse this error.
About 7\% of students computed the net force outside the numerical calculation loop, essentially making this force constant in time. 
Given students' unfamiliarity with general central force interactions, it would not be surprising if students who treated the central force outside the loop believed their solutions were correct.

\section{Closing Remarks}\label{sec:closing}

Students can develop the skills necessary to predict the motion of a variety of physical systems in large introductory physics courses. 
After solving a suite of computational homework problems, most students ($\sim 60$\%) were able to model the motion of a novel problem successfully.
In our work, we discovered that most students who were unsuccessful encountered challenges when calculating the net force acting on the object in the motion prediction algorithm.
By contrast, there were fewer students whose primary challenge was identifying and assigning variables. 
We have limited the development of our students' computational skill set to contextualizing a word problem into a programming task, identifying and updating input variables and applying a motion prediction algorithm. 
We believe that further development of our homework problems and other novel deployments could broaden the scope of the skills students develop.

The results from this work indicate that instructional efforts should be focused not only on correcting procedural mistakes but also on developing students' qualitative habits of mind. 
Students would be best served by learning the practice of debugging. 
Here, debugging includes identifying syntax errors and, more importantly, performing the type of qualitative analysis that is typically taught for solving analytic problems. 
Students who could synthesize their analytic and computational skills would be better prepared to solve the open-ended problems they will face in their future work. 

It is the goal of many reforms in physics education to develop students into flexible problem solvers while exploring the practice of science. 
Teaching computational modeling alongside physics provides support for that effort. 
Students learn the tools for doing science while developing a qualitative understanding of physical systems, exploring the generality of physics principles and learning broadly applicable problem solving methods in computation. 


\begin{theacknowledgments}
This work was supported by National Science Foundation's Division of Undergraduate Education (DUE-0618519 and DUE-0942076).
\end{theacknowledgments}



\bibliographystyle{aipproc}   

\bibliography{vp.bib}

\IfFileExists{\jobname.bbl}{}
 {\typeout{}
  \typeout{******************************************}
  \typeout{** Please run "bibtex \jobname" to optain}
  \typeout{** the bibliography and then re-run LaTeX}
  \typeout{** twice to fix the references!}
  \typeout{******************************************}
  \typeout{}
 }

\end{document}